\documentclass{PoS}
\usepackage{booktabs}
\usepackage{amsmath}
\usepackage{bm}
\usepackage{subcaption}

\title{Bottomonium spectrum at finite temperature}

\ShortTitle{Bottomonium spectrum at finite temperature}

\author{\speaker{Tim Harris} and Sin\'ead M. Ryan\\
        School of Mathematics, Trinity College, Dublin 2, Ireland\\
        E-mail: \email{tharris@tcd.ie}, \email{ryan@maths.tcd.ie}}

\author{Gert Aarts and Chris Allton\\
        Department of Physics, College of Science, Swansea University, Swansea, United Kingdom\\
        E-mail: \email{g.aarts@swan.ac.uk}, \email{c.allton@swan.ac.uk}}
\author{Seyong Kim\\
        Department of Physics, Sejong University, Seoul 143-747, Korea\\
        E-mail: \email{skim@sejong.ac.kr}}
\author{Maria Paola Lombardo\\
        INFN-Laboratori Nazionali di Frascati, I-00044, Frascati (RM) Italy\\
        E-mail: \email{mariapaola.Lombardo@lnf.infn.it}}
\author{Jon-Ivar Skullerud\\
        Department of Mathematical Physics, National University of Ireland Maynooth, Maynooth, County Kildare, Ireland\\
        E-mail: \email{jonivar@thphys.nuim.ie}}

        \abstract{We investigate the modification of S and P wave states in the bottomonium spectrum above and below the deconfinement crossover temperature through their spectral functions obtained from the maximum entropy method.
       Anisotropic ensembles with $N_f=2+1$ Wilson clover fermions with tadpole improvement are used while the bottom quark is treated with an improved non-relativistic action.
        We observe significant modifications of the P wave $\chi_{b1}$ ground state directly above the crossover temperature, $T_c$, while the S wave $\Upsilon$ ground state remains relatively unchanged up to temperatures of almost $2T_c$.
        This evidence supports earlier conclusions from our {\sc fastsum} collaboration of the immediate dissociation of the P wave states above $T_c$ and the survival of the S wave ground states up to $2T_c$.
        }

\FullConference{31st International Symposium on Lattice Field Theory LATTICE 2013\\
                 July 29 -- August 3, 2013\\
                 Mainz, Germany}

\begin{document}

\section{Probing the quark-gluon plasma with heavy quarks}
The in-medium behaviour of heavy quarkonium states offers us a valuable tool to understand the strongly-coupled plasma presumed to exist above the deconfinement crossover temperature, $T_c$,~\cite{Brambilla:2010cs,Satz:2013ama}.
The suppression of the yield of these states in nuclear collisions relative to hadronic ones can provide a signal for the formation of the deconfined phase and the temperatures reached.
Bottomonium ought to provide a cleaner probe of the hot medium than charmonium as there are fewer effects competing with the suppression of the yield~\cite{Rapp:2008tf}, while comparisons with effective field theories should be more favourable for bottom quarks.

A reliable ab initio calculation of spectral functions is desirable as these, for example in the vector channel, can be directly related to the corresponding dilepton production rate and easily compared with experimental results.
Using the lattice formulation of non-relativistic QCD (NRQCD) the only required separation of scales is $m_b\gg T$, where $m_b$ is the heavy quark mass and $T$ is the temperature.
The calculation of the spectral functions in the bottomonium system was the subject of earlier work by this collaboration which suggested dissociation of the P wave ground states immediately above $T_c$~\cite{Aarts:2010ek,Aarts:2013kaa} and survival of the S wave ground states up to at least $2T_c$~\cite{Aarts:2010ek,Aarts:2011sm}.

\begin{table}[h]
    \centering
    \begin{tabular}{r*{8}{c}}
        \toprule
        $N_s$            & 16          & 24   & 24   & 24   & 24   & 24   & 24   & 24   \\
        $N_\tau$         & 128         & 40   & 36   & 32   & 28   & 24   & 20   & 16   \\
        \midrule
        $T/T_c$          & --          & 0.76 & 0.84 & 0.95 & 1.09 & 1.27 & 1.52 & 1.90 \\
        $T$ (MeV)        & --          & 141  & 156  & 176  & 201  & 235  & 281  & 352  \\
        $N_\mathrm{cfg}$ & 499         & 502  & 503  & 998  & 1001 & 1002 & 1000 & 1042 \\
        \bottomrule
    \end{tabular}
    \caption{Summary of the ensembles used in this work.
        Each ensemble has the following parameters: lattice spacing $a_s=0.1227(8)$ fm, anisotropy $\xi\equiv a_s/a_\tau=3.5$ and pion mass $M_\pi\approx400$ MeV~\cite{Edwards:2008ja}.
        The crossover temperature is determined from the renormalized Polyakov loop~\cite{Allton:2013aa}.}
    \label{tab:ensem}
\end{table}
In this work we employ anisotropic ensembles with $N_f=2+1$ flavours of tadpole-improved Wilson clover quark which represent multiple improvements over ensembles used in previous studies by this collaboration~\cite{Aarts:2010ek}.
The tuning of the zero temperature ensembles was completed by the Hadron Spectrum Collaboration~\cite{Edwards:2008ja}.
In addition to the inclusion of the dynamical strange quark, there is a finer spatial lattice spacing and a pion mass closer to the physical value compared with ensembles used in earlier studies, see table~\ref{tab:ensem} for details. 
%The ensemble details are available in table~\ref{tab:ensem}.
Anisotropic lattice spacings allow high temperatures to be achieved without sacrificing too many temporal sites or incurring significant computational cost.
Furthermore, the fixed-scale approach reduces the number of zero temperature simulations required to tune the lattice parameters such as the heavy quark mass.

\section{NRQCD at zero and non-zero temperature}
The NRQCD quark propagators solve an initial value problem which is evident from the form of the heavy quark action~\cite{Lepage:1992tx}:
\begin{align}
    S=a_s^3\sum_{x\in\Lambda}{\psi^\dagger}(x)\left(\psi(x)-
    \left[1-\frac{a_\tau H_0|_{\tau+a_\tau}}{2}\right]U_\tau^\dagger(x)
    \left[1-\frac{a_\tau H_0|_{\tau}}{2}\right]\left(1-{a_\tau\delta H}\right)  
    \psi(x-a_\tau \bm e_\tau)\right) 
%    G(\bm x, \tau+a_\tau)=K(\tau+a_\tau)G(\bm x,\tau)
    \label{eq:action}
\end{align}
where $\psi(x)$ is the heavy quark field, $U_\tau(x)$ the temporal gauge links and $\bm e_\tau$ the temporal unit vector.
The leading order Hamiltonian, $H_0$, and covariant finite difference operators are given in terms of the usual forward and backward ones
\begin{align}
    H_0=-\frac{\Delta^{(2)}}{2m_b},
        \qquad\Delta^{(2n)}
        =\sum_{i=1}^3\left(\nabla_i^+\nabla_i^-\right)^{n}.
\end{align}
The improved NRQCD action includes relative $O(v^2)$ relativistic corrections to the leading term and relative $O(v^4)$ spin-dependent terms, where $v\sim|\bm p|/m_b$ is the quark velocity in the bottomonium rest frame, the power counting parameter in this effective theory.
Furthermore, corrections are included to remove $O(a_\tau)$ errors in the evolution and $O(a_s^2)$ errors in the leading order piece
\begin{align}
    \label{eq:deltaH}
    \begin{split}
        \delta H=-\frac{\left(\Delta^{(2)}\right)^2}{8m_b^3}
                +\frac{ig_0}{8m_b^2}\left(\bm\nabla^\pm\cdot \bm E
                - \bm E\cdot\bm\nabla^\pm\right)
            &-\frac{g_0}{8m_b^2}\bm\sigma\cdot\left(\bm\nabla^\pm\times \bm E
                - \bm E\times\bm\nabla^\pm\right)-\frac{g_0}{2m_b}\bm \sigma\cdot\bm B\\
            &+\frac{a_s^2\Delta^{(4)}}{24m_b}
                -\frac{a_\tau\left(\Delta^{(2)}\right)^2}{16 m_b^2}.
    \end{split}
\end{align}
Unimproved fields are defined through the usual clover definition of the field strength tensor, while tadpole improvement is implicit everywhere by dividing all spatial links by the fourth root of the spatial plaquette.

Since the rest mass term in the heavy quark dispersion relation can be removed by a field redefinition only energy splittings are meaningful.
%This entails tuning the heavy quark mass via a hadronic dispersion relation.
We tune the spin-averaged 1S kinetic mass, $M_2(\overline{1\mathrm S})\equiv(M_2(\eta_b)+3M_2(\Upsilon))/4$ to its experimental value to mitigate the systematic error caused by the incorrect determination of the hyperfine splitting~\cite{Dowdall:2011wh}.
When comparing spectral quantities the $\Upsilon$ mass is fixed to its experimental value to set the absolute energy.

The NRQCD formulation has certain advantages when interpreting the modification of the spectrum at finite temperature.
As the heavy quark propagators satisfy an initial value problem they do not obey the thermal boundary conditions like the other fields.
Thermal effects are introduced only through interaction with the hot medium.
Hadronic correlation functions are asymmetric in time and the heavy quarks are not in thermal equilibrium with the medium.
This asymmetry is reflected in the temperature independent kernel, $K(\tau,\omega)=e^{-\omega\tau}$, which appears in the integral representation of the hadronic correlation function
\begin{align}
    G(\tau;T)=\int_{\omega_{\rm min}}^{\omega_{\rm max}} \frac{d\omega}{2\pi}\,K(\tau,\omega)\rho(\omega;T).
    \label{eq:intrep}
\end{align}
Thermal modifications to the correlator arise only from the temperature dependence of the spectral function.
Furthermore, there is no zero mode type contribution at any temperature which usually gives rise to a constant contribution to correlation functions.
%Since no transport information is encoded in the correlation functions all thermal effects arise from the modification of the spectrum.

The spectral functions are obtained from eq.~(\ref{eq:intrep}) by using the maximum entropy method (MEM)~\cite{Asakawa:2000tr} with Bryan's algorithm.
This Bayesian method ensures the solution for the spectral function, $\rho(\omega)$, is unique although independence of the prior information must be demonstrated.

\begin{figure}[htpb]
    \centering
        \begin{subfigure}[t]{0.4\textwidth}
            \includegraphics[width=\textwidth]{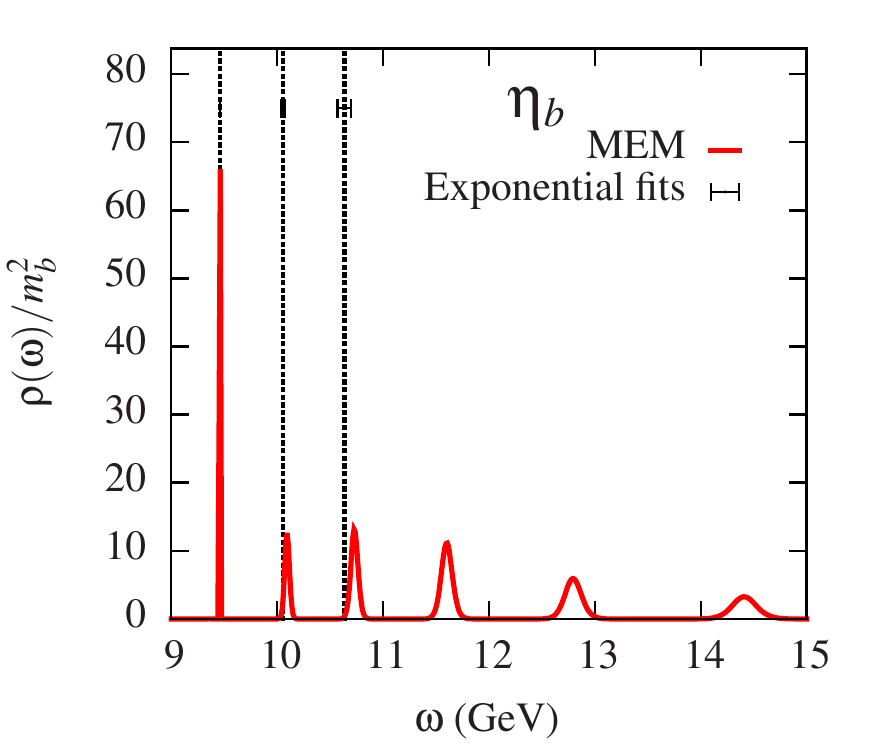}
        \end{subfigure}%
        \begin{subfigure}[t]{0.4\textwidth}
            \includegraphics[width=\textwidth]{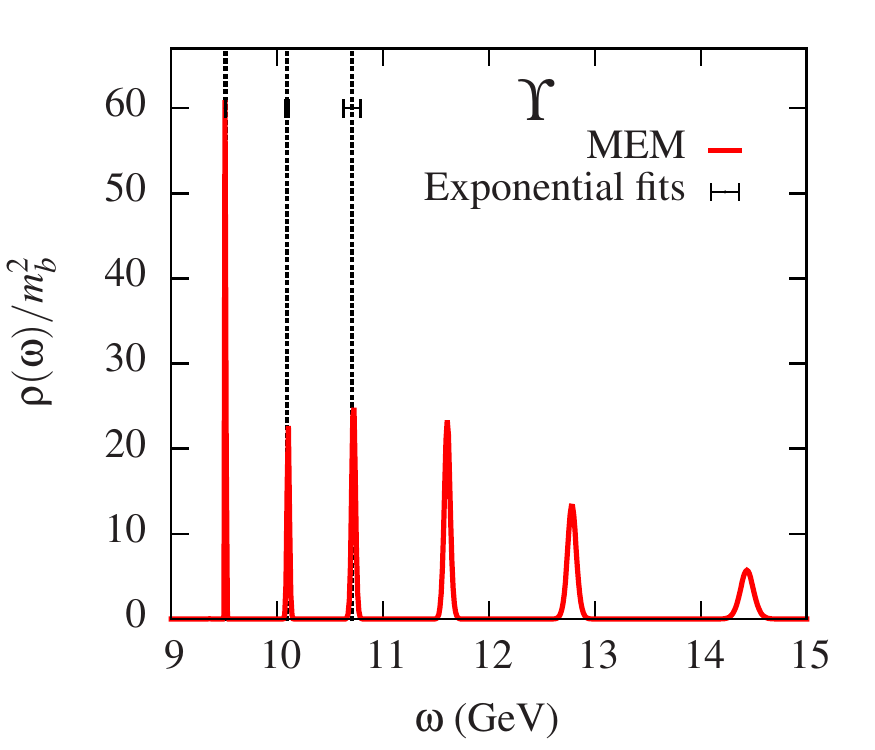}
        \end{subfigure}
        \caption{Reconstructed spectral function at the lowest temperature in the $\eta_b$ (left) and $\Upsilon$ (right) channels, with energies extracted from multi-exponential fits displayed with dashed lines.}
    \label{fig:rho_zeroT}
\end{figure}

The asymmetric correlation functions allow access to larger temporal separations than for relativistic quarks which ameliorates the reconstruction of the spectral function.
The zero temperature spectral functions are shown in figure~\ref{fig:rho_zeroT} in the S wave $\eta_b$ (left) and $\Upsilon$ (right) channels.
As many as six peaks can be seen, corresponding to bound states.
The agreement of the peak positions with the energies extracted from multi-exponential fits shown with dashed lines, obtained with the \textsc{CorrFitter} package~\cite{Hornbostel:2011hu}, is good.
\section{Thermal modification of the bottomonium spectrum}
\begin{figure}[htpb]
    \centering
        \begin{subfigure}[t]{0.5\textwidth}
            \includegraphics[width=\textwidth]{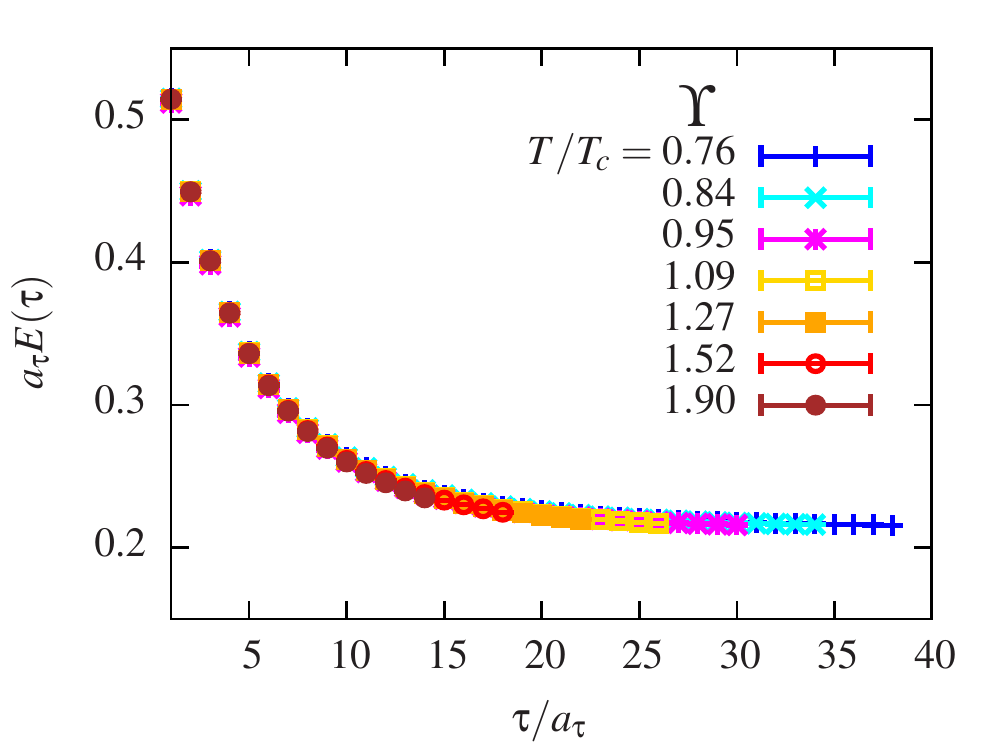}
        \end{subfigure}%
        \begin{subfigure}[t]{0.5\textwidth}
            \includegraphics[width=\textwidth]{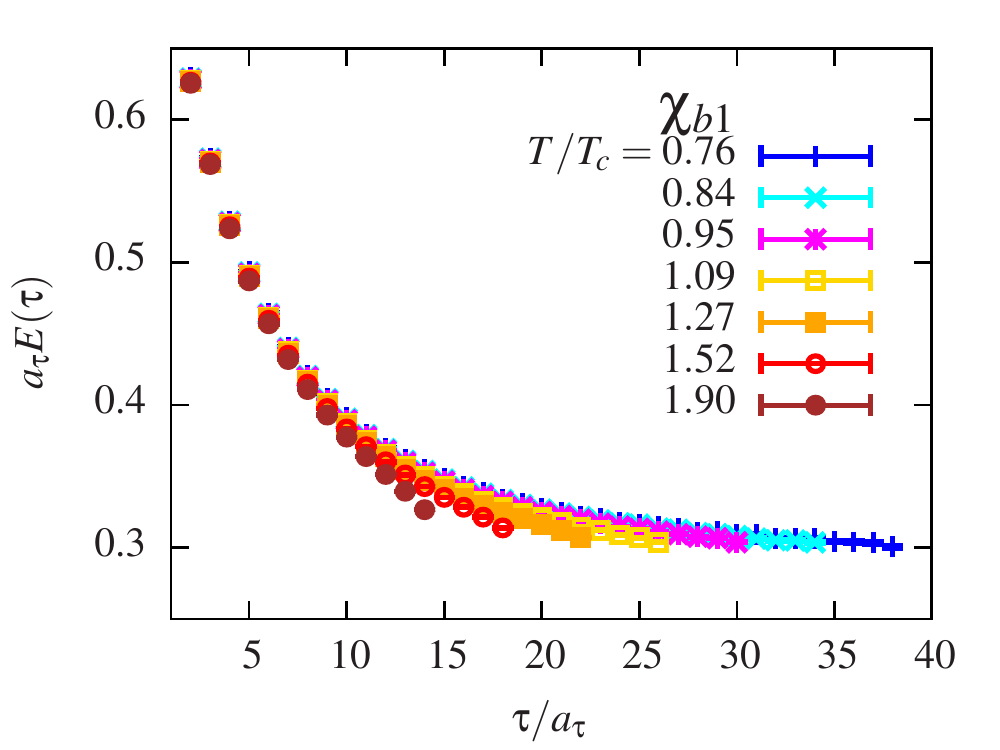}
        \end{subfigure}
        \caption{Temperature dependence of the effective mass in the $\Upsilon$ (left) and $\chi_{b1}$ (right) channels.}
    \label{fig:meff_tdep}
\end{figure}

Figure~\ref{fig:meff_tdep} illustrates the temperature dependence of the effective mass in the S wave $\Upsilon$ (left) and P wave $\chi_{b1}$ (right) channels.
The $\chi_{b1}$ channel displays significant modifications while the $\Upsilon$ channel appears relatively unchanged as the temperature is varied.
We can gain some insight into the temperature dependence by comparing with the predictions for free quarks, valid in the high-temperature non-interacting limit, where the continuum spectral functions are known~\cite{Burnier:2007qm}
\begin{align}
    \rho_{\textrm{free}}(\omega)\propto(\omega-\omega_0)^{\alpha}\,\Theta(\omega-\omega_0)
        \quad\Rightarrow\quad
    G_{\textrm{free}}(\tau)\propto \frac{e^{-\omega_0\tau}}{\tau^{\alpha+1}},
    \label{eq:rhofree}
\end{align}
where $\alpha=1/2$ for S waves and $\alpha=3/2$ for P waves.
A non-zero threshold $\omega_0$ is included to account for the unknown additive shift in the energies.
It is in general temperature dependent and depends on the lattice parameters.
This form for the correlation functions leads to an effective mass with a power law behaviour at asymptotically large temporal separations
\begin{align}
    E(\tau) = -\frac{1}{G(\tau)}\frac{dG(\tau)}{d\tau}
    \quad\stackrel{G=G_\mathrm{free}}{\longrightarrow}\quad
    \omega_0+\frac{\alpha+1}{\tau}.
    \label{eq:effmass}
\end{align}
%As can be seen in figure~\ref{fig:meff_tdep}, the P waves exhibit deviation from the plateaux observed in the S wave channel.
Although at high temperatures the P wave channel deviates from the plateaux exhibited at low temperatures, the effect of a larger threshold compared with the earlier study~\cite{Aarts:2010ek} may make the effect appear less pronounced. 
%Compared with the effect seen in the earlier work in the P wave channel, the effect is less pronounced at similar temperatures.
%A larger threshold would account for a less dramatic effect seen in the effective energies, and indeed in the spectrum calculation the threshold is identified as being larger than in the previous study.
%

\section{Heavy quark mass dependence}
Along with Landau damping, whose role has been highlighted in effective field theory studies~\cite{Burnier:2007qm}, colour-Debye screening plays a significant part in the dissociation of heavy quarkonium bound states in the plasma.
%Although effective field theory analyses have emphasised the role of the finite width of heavy quarkonium states in the plasma~\cite{Burnier:2007qm}, colour-Debye screening plays a significant part in the dissociation of bound states.
When the Debye screening length is comparable with the binding radius of the state, colour-screening of the charges will lead to its break-up.
In potential models we expect the binding radius to depend inversely on the quark mass, $R\sim(m_bv)^{-1}$, so that heavier, more tightly bound states will experience less screening and survive to greater temperatures in the plasma.
\begin{figure}[htpb]
    \centering
        \begin{subfigure}[t]{0.5\textwidth}
            \includegraphics[width=\textwidth]{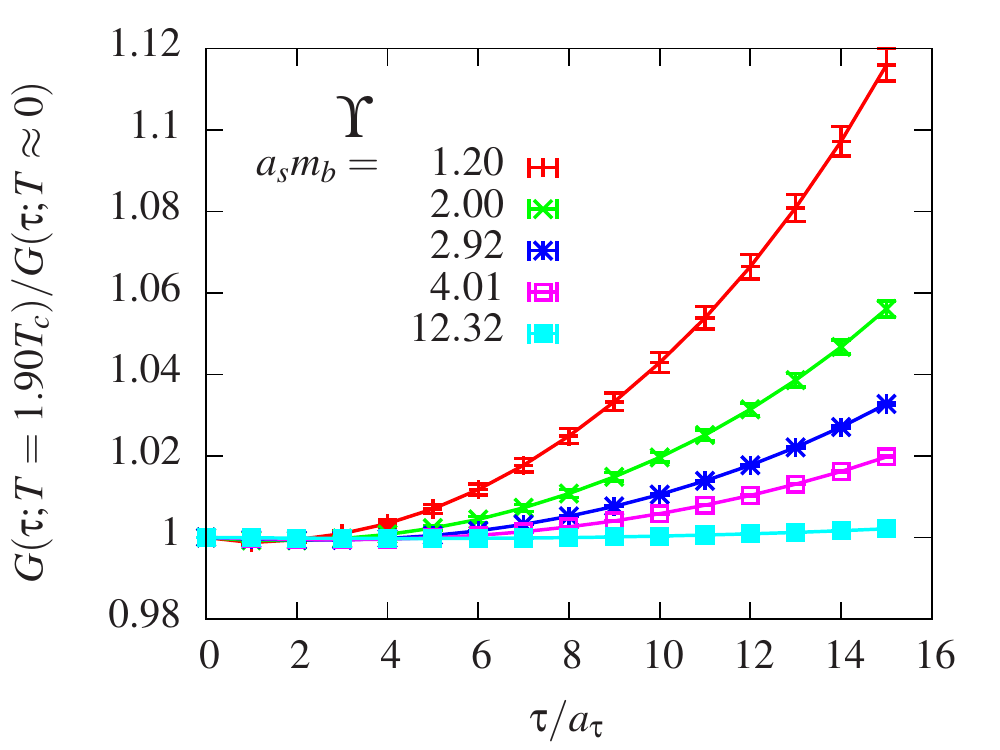}
        \end{subfigure}%
        \begin{subfigure}[t]{0.5\textwidth}
            \includegraphics[width=\textwidth]{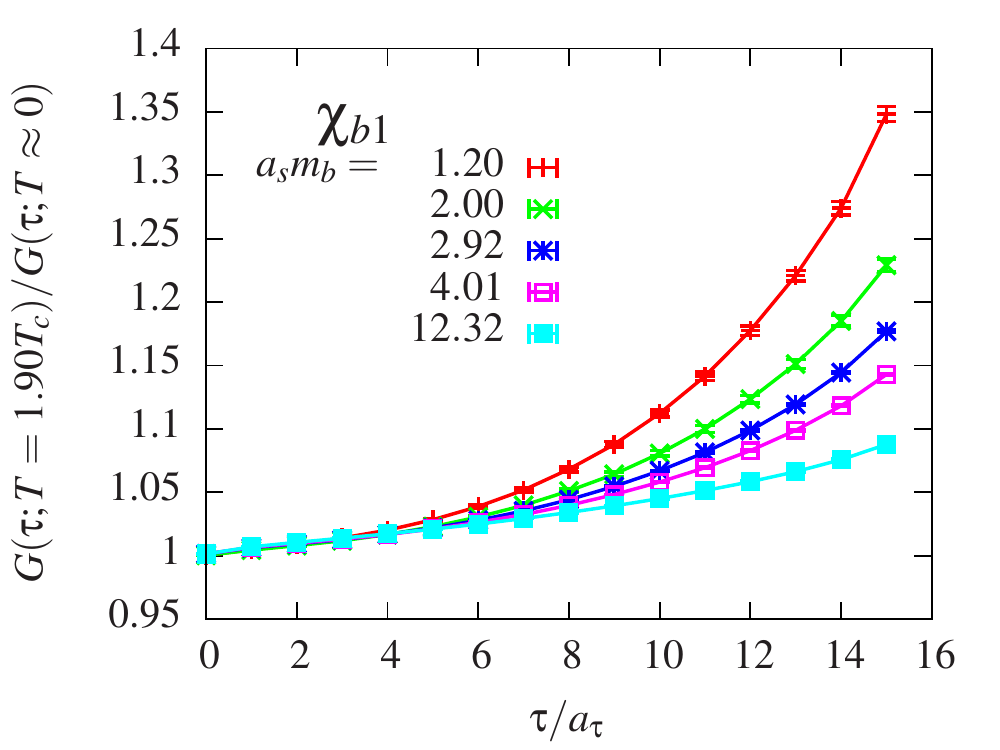}
        \end{subfigure}
        \caption{Dependence of the modification in the correlation function in the $\Upsilon$ (left) and $\chi_{b1}$ channels at the highest accessible temperature, $T=1.90T_c$. The physical heavy quark mass is $a_sm_b=2.92$.}
    \label{fig:m0dep}
\end{figure}
Figure~\ref{fig:m0dep} shows the mass dependence of the modification to the correlation function at a fixed temperature, $T/T_c=1.90$.
In line with these expectations, the heavier states display less modification than the lighter ones at large time separations. The $\chi_{b1}$ channel (right) shows greater enhancement than the $\Upsilon$ channel (left) at each value of the heavy quark mass.
Note that at the tuned physical heavy quark mass, $a_sm_b=2.92$, the correlation function in the $\Upsilon$ channel sees only about 3\% enhancement at large separations, while the $\chi_{b1}$ channel shows almost 20\% enhancement with respect to the zero temperature one.

\section{Bottomonium spectral functions above and below $T_c$}
\begin{figure}[htpb]
    \centering
        \begin{subfigure}[t]{0.82\textwidth}
            \includegraphics[width=\textwidth]{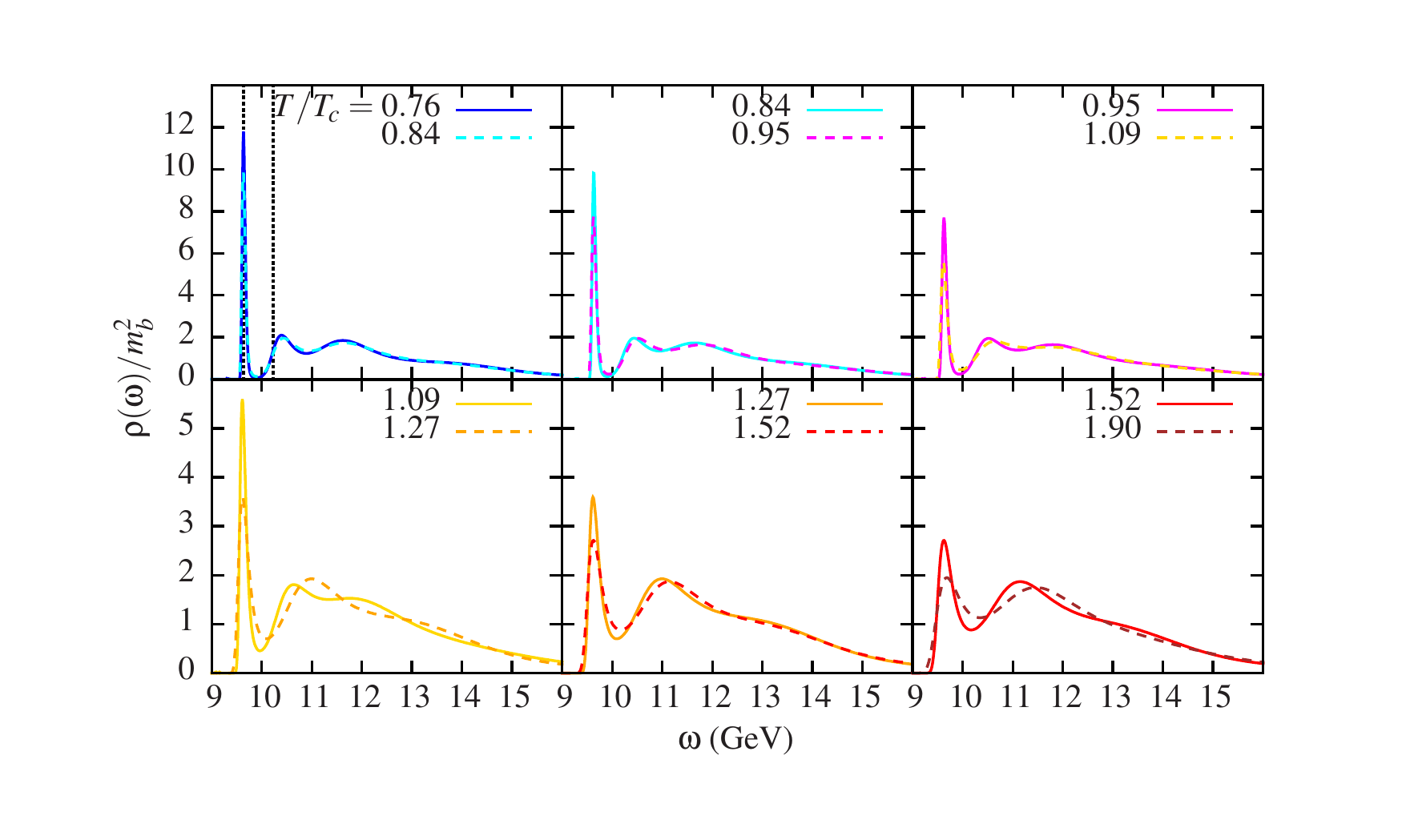}
        \end{subfigure}
        \vskip -2em
        \begin{subfigure}[t]{0.82\textwidth}
            \includegraphics[width=\textwidth]{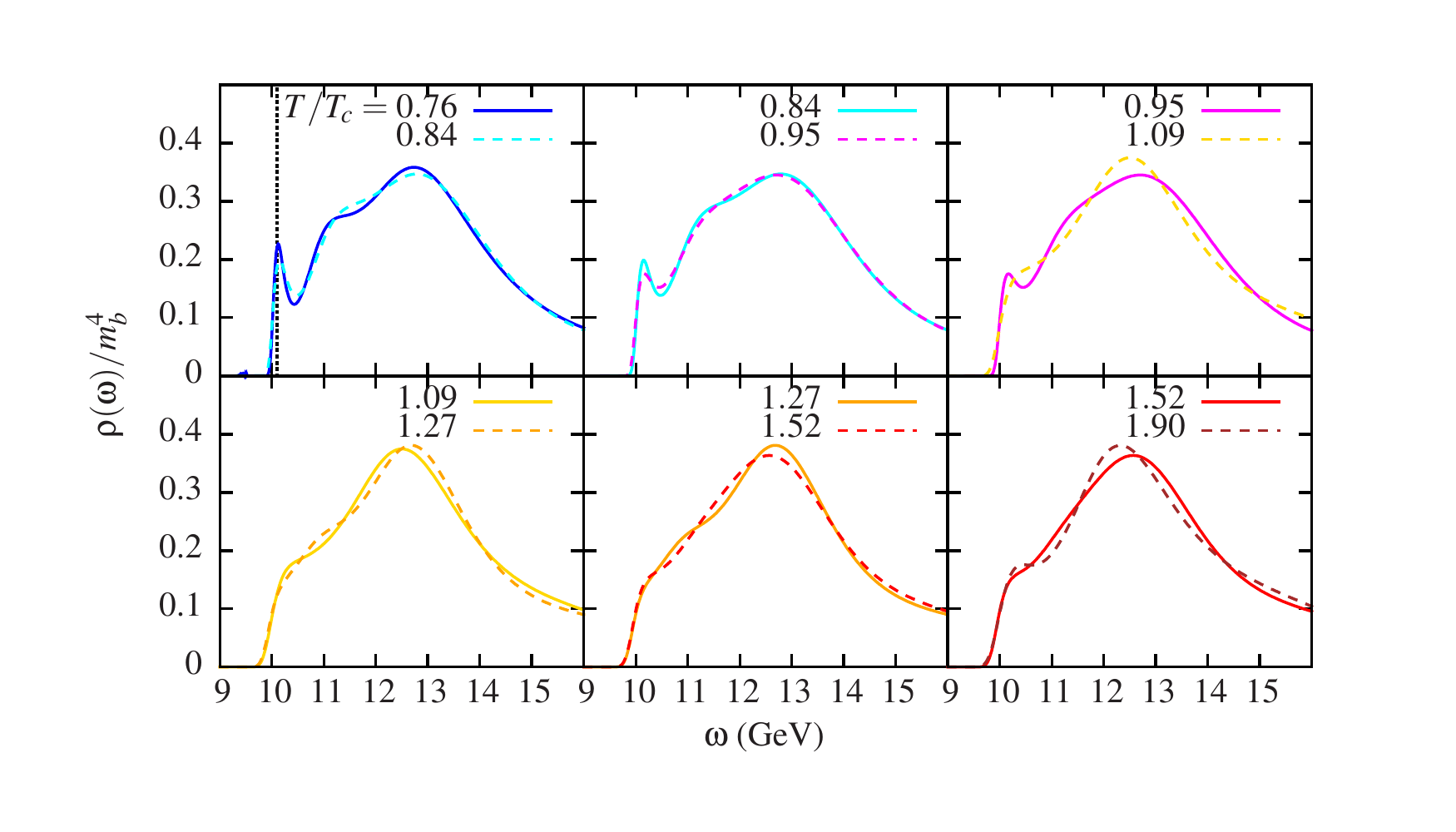}
        \end{subfigure}
        \caption{Temperature dependence of the spectral functions in the $\Upsilon$ (upper figure) and $\chi_{b1}$ (lower figure) channels.
        Each panel displays only two neighbouring temperatures.
        Note the different ordinate scale between the upper and lower panels in the upper figure.}
    \label{fig:rho_Tdep}
\end{figure}

We show the spectral functions and their evolution with temperature in figure~\ref{fig:rho_Tdep}.
In both the $\Upsilon$ channel (upper figure) and $\chi_{b1}$ channel (lower figure) the ground state peaks are evident at temperatures below $T_c$ and coincide with the energy extracted from exponential fits to the zero temperature correlators.

In the $\Upsilon$ channel a peak close to the first excited state, $\Upsilon(2\mathrm S)$, is also apparent at temperatures below $T_c$, while the ground state peak remains visible at all accessible temperatures.
We understand this to indicate the survival of the $\Upsilon$ ground state to temperatures up to $1.90T_c$.
The modification of the $\Upsilon(2\mathrm S)$ is less straightforward to interpret as the peak seems to merge with lattice artefacts above $T_c$ as it shifts and broadens.

The $\chi_{b1}$ ground state can no longer be seen immediately above $T_c$ which suggests that this state might dissociate immediately in the deconfined phase.
We note that the support of the spectral function at high temperatures is comparable with the support of the free lattice spectral function which is not shown in this figure.

\section{Conclusions}

We have calculated bottomonium spectral functions using MEM at temperatures above and below $T_c$ on a new generation of anisotropic $N_f=2+1$ Wilson clover ensembles.
Our results are compatible with conclusions from previous studies~\cite{Aarts:2010ek,Aarts:2011sm,Aarts:2013kaa} and have been obtained in the same temperature range as experiments so we hope they help with the interpretation of the experimental spectra~\cite{Chatrchyan:2012lxa}.
%The results are compatible with conclusions from previous studies~\cite{Aarts:2010ek,Aarts:2011sm,Aarts:2013kaa} and appear to be consistent with recent experimental results in the S wave $\Upsilon$ channel~\cite{Chatrchyan:2012lxa}.
A discussion of the systematic checks of the MEM method will be presented in future along with a comparison with the free lattice spectral functions.

\paragraph{Acknowledgements}
This work is undertaken as part of the UKQCD collaboration and the STFC funded DiRAC Facility.
We acknowledge the PRACE Grants 2011040469 and Pra05 1129, the Initial Training Network STRONGnet, Science Foundation Ireland, the Trinity Centre for High-Performance Computing, the Irish Centre for High-End Computing, STFC, the Wolfson Foundation, the Royal Society and the Leverhulme Trust for support.
We thank D.~K. Sinclair for use of his NRQCD code and A. Rothkopf for stimulating discussions about the Bayesian reconstruction of the spectral functions.

\end{document}